\documentclass[lettersize,journal]{IEEEtran}
\usepackage{amsmath,amsfonts}
\usepackage{algorithmic}
\usepackage{algorithm}
\usepackage{array}
\usepackage[caption=false,font=normalsize,labelfont=sf,textfont=sf]{subfig}
\usepackage{textcomp}
\usepackage{stfloats}
\usepackage{url}
\usepackage{verbatim}
\usepackage{graphicx}
\usepackage{cite}
\usepackage{hyperref}

\usepackage{enumitem}

\DeclareMathOperator*{\argmax}{argmax}
\begin{document}

\onecolumn 

\begin{description}

\item[\textbf{Citation}] R. Benkert, M. Prabhushankar, and G. AlRegib, “Effective Data Selection for Seismic Interpretation
through Disagreement,” in IEEE Transactions on Geoscience and Remote Sensing, April 2024 \\


\item[\textbf{Review}]
{Date of submission: 21 July 2023\\
Date of 1st revision: 11 September 2023\\
Date of 2nd revision: 29 January 2024\\
Date of acceptance: 29 April 2024
} \\


\item[\textbf{Bib}]{@article\{benkert2024effective,\\
  title=\{Effective Data Selection for Seismic Interpretation
through Disagreement\},\\
  author=\{Benkert, Ryan and Prabhushankar, Mohit, and AlRegib\},\\
  journal=\{IEEE Transactions on Geoscience and Remote Sensing\},\\
  year=\{2024\},\\
  publisher=\{IEEE\}\} 
} \\


\item[\textbf{Copyright}]{© 2024 IEEE.  Personal use of this material is permitted.  Permission from IEEE must be obtained for all other uses, in any current or future media, including reprinting/republishing this material for advertising or promotional purposes, creating new collective works, for resale or redistribution to servers or lists, or reuse of any copyrighted component of this work in other works.}
\\
\item[\textbf{Contact}]{\href{mailto:rbenkert3@gatech.edu}{rbenkert3@gatech.edu}  OR \href{mailto:alregib@gatech.edu}{alregib@gatech.edu}\\ \url{http://ghassanalregib.info/} \\ }

\end{description}

\thispagestyle{empty}
\newpage
\clearpage
\setcounter{page}{1}

\twocolumn

\title{Effective Data Selection for Seismic Interpretation through Disagreement}

\author{{Ryan Benkert,~\IEEEmembership{Student Member,~IEEE,}, Mohit Prabhushankar,~\IEEEmembership{Member,~IEEE,}, \\ and AlRegib, Ghassan,~\IEEEmembership{Fellow,~IEEE,}}
\thanks{The authors are with the School of Electrical and Computer Engineering, College of Engineering, Georgia Institute of Technology, Atlanta, GA 30332 USA (e-mail: rbenkert3@gatech.edu; mohit.p@gatech.edu; alregib@gatech.edu)}
}

\markboth{Accepted to IEEE Transactions on Geoscience and Remote Sensing 2024}
{Ryan Benkert et. al.: Effective Data Selection for Seismic Interpretation
through Disagreement}


\maketitle

\begin{abstract}


This paper presents a discussion on data selection for deep learning in the field of seismic interpretation. In order to achieve a robust generalization to the target volume, it is crucial to identify the specific samples are the most informative to the training process. The selection of the training set from a target volume is a critical factor in determining the effectiveness of the deep learning algorithm for interpreting seismic volumes. This paper proposes the inclusion of interpretation disagreement as a valuable and intuitive factor in the process of selecting training sets. The development of a novel data selection framework is inspired by established practices in seismic interpretation. The framework we have developed utilizes representation shifts to effectively model interpretation disagreement within neural networks. Additionally, it incorporates the disagreement measure to enhance attention towards geologically interesting regions throughout the data selection workflow. By combining this approach with active learning, a well-known machine learning paradigm for data selection, we arrive at a comprehensive and innovative framework for training set selection in seismic interpretation. In addition, we offer a specific implementation of our proposed framework, which we have named \texttt{ATLAS}. This implementation serves as a means for data selection. In this study, we present the results of our comprehensive experiments, which clearly indicate that \texttt{ATLAS} consistently surpasses traditional active learning frameworks in the field of seismic interpretation. Our findings reveal that \texttt{ATLAS} achieves improvements of up to 12\% in mean intersection-over-union.

\end{abstract}

\section{Introduction}
Deep learning models have become the preferred method in seismic interpretation, bringing about a revolutionary change by reducing the time and effort needed for experts to annotate seismic volumes \cite{alregib2018subsurface}.
Instead of geophysicists annotating seismic volumes for weeks \cite{wang2018successful}, deep models can infer similar information within a matter of hours. The success of machine learning algorithms relies heavily on sufficient quantities of training data and a well-constructed optimization process. When provided with accurate training data, model representations effectively capture geological information and mirror the manual interpretation process. 
However, when faced with scarce training sets, model representations deteriorate, leading to unpredictable behavior and semantically incoherent outputs \cite{spm}.  


Unfortunately, scarcity and label inaccuracies are commonplace in seismic datasets, posing significant challenges for the deployment of deep learning models. For seismic interpretation, training labels require expert annotation and can substantially differ depending on the available information to the interpreter. As a result, labeled training sets are frequently a small subset of the entire seismic volume and the algorithmic performance is heavily influenced by location and content of the individual training sections. Therefore, selecting the appropriate training volume is critical for the successful deployment of deep learning models. 

Within the context of machine learning, a common mechanism for limited data settings is active learning. Active learning is a machine learning paradigm where the deep learning algorithm selects an appropriate training set that maximizes test performance \cite{cohn1996active}. 
Traditionally, active learning is an iterative process, where the model is trained on an initial set of annotated training samples and actively selects additional samples in subsequent rounds. While active learning has proven successful in a wide range of natural image applications \cite{benkert2022forgetful, ash2019deep, houlsby2011bayesian}, its direct applicability to seismic data acquisition workflows is not trivial. Specifically, active learning algorithms must assign an importance score to entire seismic sections instead of individual pixels. In seismic data, regions of geological interest constitute complex structures and underrepresented classes that frequently constitute minor portions of the seimsic section. Consequently, blindly applying active learning algorithms can result the selection of arbitrary sections as relevant geological areas are neglected when combined into a single section score value. In other words, we argue that successful data selection in seismic interpretation requires \emph{spatial awareness} to informative geological regions. 



\begin{figure*}
    \begin{center}
        \includegraphics[scale=0.35]{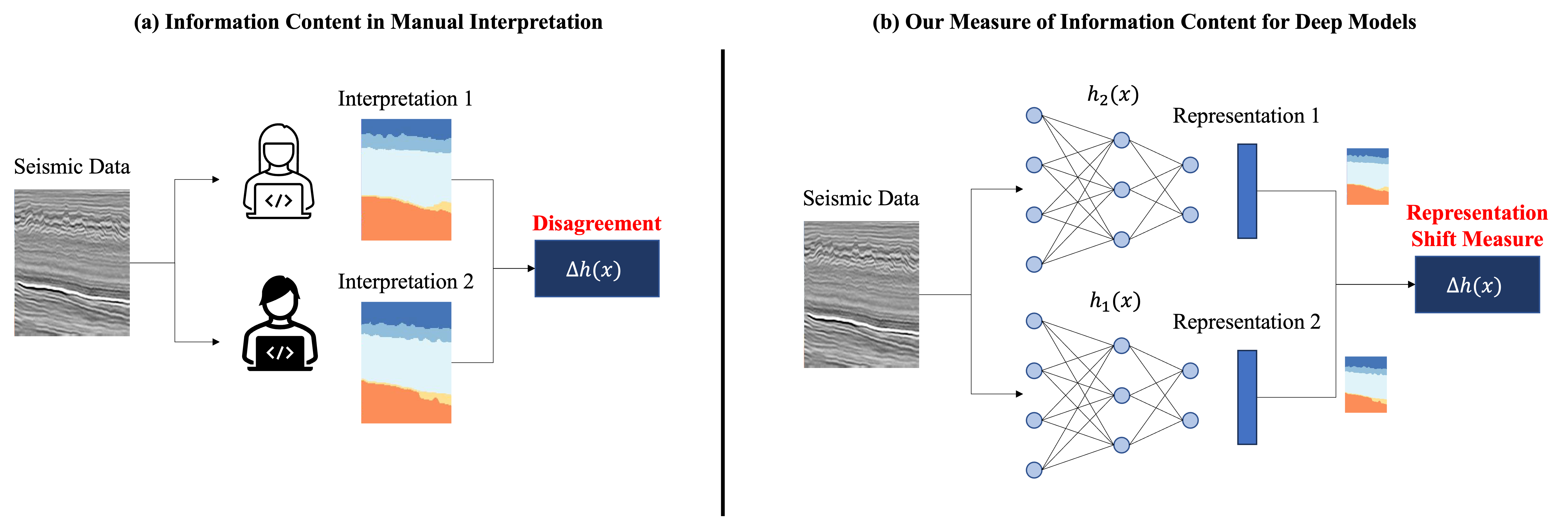}
    \end{center}
    \caption{High-level overview of our measure of information content in seismic interpretation. \textbf{a)} information content for manual interpretation. Two different interpreters annotate the same seismic data section. Information content is measured by the disagreement between both interpretations $\Delta h$. \textbf{b)} our measure of information content with deep neural networks. Information content is measured by the disagreement or shift between representations of the same seismic section $\Delta h$.}
    \label{fig:spatially-aware-wf}
\end{figure*}

This paper concerns data selection for the application of seismic interpretation. Mainly, the paper formulates a spatially aware framework to select the most informative seismic sections for deep learning algorithms. From a geophysical standpoint, a common definition of informative content would entail label ambiguity. In fact, it is common practice for different interpreters to disagree on annotations and significantly alter interpretations when additional information becomes available. In Figure~\ref{fig:spatially-aware-wf}a, we show a toy example of information content for manual interpretation. Information is measured by evaluating the disagreement ($\Delta h$) between two separate interpretations of the same section. In this paper, we argue that incorporating interpretation disagreement within the selection of the training volume is both intuitive and beneficial in terms of generalization performance. For neural networks, interpretations are based on the mathematical representations of the individual data points within a seismic section. As a result, disagreement between neural networks is measured in terms of difference or \emph{shift} between the representations. Figure~\ref{fig:spatially-aware-wf}b, gives a high level overview of the measurement. In order to imitate the manual interpretation process, we leverage two separate networks and measure disagreement through the shift between the mathematical representations (denoted as $\Delta h$). We formalize the concept mathematically in Section~\ref{sec:representation-shifts} and provide an implementation of representation shift measurements in Section~\ref{sec:atlas-methodology}. 



Within our selection framework, disagreement plays an instrumental role for identifying regions of geological interest spatially within a seismic section. Specifically, we measure the representation shift between two separate neural networks trained on labeled training data. Subsequently, we restrict the input image to regions of high disagreement by masking with a representation shift measure. The masked image is then fed to the active learning algorithm. Our algorithm represents a strong contrast to conventional active learning where the input image is fed to the algorithm directly. We provide a formalization of our framework in Section~\ref{sec:spatially-aware-active-learning}. In Section~\ref{sec:atlas-methodology}, we provide an instance of our framework, \texttt{ATLAS}, which significantly outperforms conventional active learning pipelines with a variety of popular acquisition algorithms. We perform detailed analysis and experiments in Section~\ref{sec:results}, and report improvements up to 10\% in mean-intersection-over-union on a variety of popular active learning algorithms.

In summary, the contributions of this paper are as follows:
\begin{enumerate}
    \item We identify representation shifts as an attractive tool for data selection in seismic interpretation.
    \item We develop \texttt{ATLAS}, a plug-in algorithm for targeted data selection in seismic interpretation.
    \item We perform empirical evaluation of \texttt{ATLAS} on two seismic surveys and provide both quantitave and qualitative evidence for performance improvements up to 10\% in mean-intersection-over-union on a variety of popular active learning algorithms.
\end{enumerate}





\section{Related Work}

\subsection{Sampling Efficiency in Seismic Interpretation}
The present literature to data selection for machine learning in seismic interpretation is relatively narrow in its coverage.
Initially, deep learning models for seismic data were predominantly developed under fully supervised settings \cite{di2018patch, wu2018convolutional}. 
However, the high cost associated with annotating large volumes of seismic data presents a challenge in building large-scale models. 
Consequently, researchers have explored alternative approaches, such as semi-supervised, self-supervised, or weakly supervised methods, to alleviate the annotation burden during model training \cite{alaudah2016weakly, alaudah2018structure, alaudah2019facies, alaudah2018learning, alfarraj2019semisupervised, babakhin2019semi, kokilepersaud2022volumetric, wang2022self, yuan2022self}. While these approaches succeed in reducing the number of required training sections, they are not interactive sampling strategies and therefore cannot be considered as active learning acquisition functions.

In addition to methodological advancements, researchers have explored different modalities and applications within seismic interpretation, each with their own distinct data requirements. 
For example, fault detection has been addressed in various studies \cite{araya2017automated, di2019improving, di2019semi, shafiq2018novel, wu2019faultseg3d, xiong2018seismic}, while the delineation of salt bodies has been investigated as well \cite{di2018deep, di2018multi, shafiq2015detection}. Classification of facies, has seen contributions from multiple researchers \cite{liu20193d, dramsch2018deep, qian2017seismic, alaudah2019facies, alaudah2019machine}. Another area of interest involves predicting rock lithology from well logs, where researchers have explored semi-supervised approaches \cite{alfarraj2019semisupervised, das2018convolutional, das2019effect, ahmadjournal}. In each case, annotation requirements are unique and result in a different annotational burden. Our work is tangential, and applicable to arbitrary settings where annotations are limited and costly.


In all the previously discussed related work, the training set was considered fixed without active acquisition of new training sections. Recently, a few research directions have explored data selection pipelines, representing the initial steps toward  active learning algorithms in seismic interpretation \cite{di2022automated, di2022automatedLE, mustafa2021man, mustafa2023active}. However, each of these approaches follows a conventional active learning workflow where informativeness is evaluated over the entire seismic section. 
The utilization of this approach for seismic data is considered unfavorable due to the underrepresentation of geophysically significant areas, such as faults and salt domes. These areas only contribute minimally to the informativeness score.
Our work enhances these approaches by limiting the active learning algorithm to areas of geophysical significance, reducing the focus on commonly occurring or well-represented structures or classes. It is important to note that our method has the capability to be utilized with any active learning acquisition function, which ultimately leads to a beneficial workflow for seismic interpretation.

\subsection{Active Learning}
Within the context of machine learning itself, data selection methods are well established with extensive research in active learning active learning \cite{dasgupta2011two, settles2009active, hanneke2014theory}. Within this context, our work strongly relates to the practical study of different data selection functions. 
In our research, we draw on the principles and methodologies of active learning, specifically focusing on the study of various methods to acquire unlabeled data. These methods play a crucial role in determining the importance ranking of unlabeled data points based on their information content. We will review several prominent active learning algorithms and differentiate them based on their respective definitions of information content.

The first significant research direction involves defining information content with generalization difficulty \cite{wang2014new, benkert2022forgetful, roth2006margin, schohn2000less}. For instance, several approaches define informativeness with softmax probabilities \cite{wang2014new, roth2006margin} where information content is related to the output logits of the network. Closely related, \cite{schohn2000less, tong2001support} consider information in the for of decision boundary proximity. Other directions \cite{gal2017deep} interpret generalization difficulty as model uncertainty and establish acquisition functions with Monte-Carlo Dropout. The second direction, defines information content based on data representation within the dataset. Here, a popular family of algorithms construct the core-set of the unlabeled data pool \cite{sener2017active, longtailcoreset}. Further, \cite{gissin2019discriminative} reformulate active learning as an adversarial training problem and \cite{logan2022decal, logan2022patient} consider prior representative information within the data. The third category considers combinations of generalization difficulty and data representation. In several cases \cite{ash2019deep, benkert2023gaussian, prabhushankarintrospective}, algorithms establish a ranking based on difficulty and sample based on representational components. \cite{batchbald} extend the difficulty based algorithm \cite{houlsby2011bayesian} to diversifying batch acquisitions. Finally, \cite{baram2004online, hsu2015active} propose "bandit-style" algorithms that interactively alternate between difficulty-based, and representational algorithms each active learning round. In this paper, we benchmark our method on difficulty based approaches due to their particular interest in seismic interpretation. In particular, seismic machine learning workflows are frequently directed towards complex structures and underrepresented classes (e.g. faults and salt domes); components that are unrepresentative of the entire volume and especially targeted by difficulty-based approaches. 

Finally, several active learning algorithms exist that target semantic segmentation explicitly \cite{siddiqui2020viewal, wu2021redal, xie2020deal}. Our work is complementary to these works along two dimensions: first, existing approaches represent separate acquisition functions instead of a plug-in framework applicable \emph{on top} of any active learning algorithm. Hence, our approach is modular and can be utilized in combination with current acquisition functions, as well as novel approaches operating on seismic data. Second, current segmentation approaches in active learning are not designed on seismic data which comes with fundamentally different challenges. For instance, \cite{siddiqui2020viewal} label individual super-pixels in lieu of the entire input image. While effective for several natural image applications, the approach is not applicable to settings where data points are labeled in their entirety; a common case in seismic data workflows.

\subsection{Deep Feature Extraction for Seismic Interpretation}
Our work further relates to deep feature extraction applied to seismic interpretation with a focus on informative representation learning. For seismic interpretation, feature extraction is frequently studied within the context of carefully constructed attributes. Attributes represent quantitative measures (attributes) of interesting geological characteristics that can be used to infer information about the subsurface \cite{chopra2005seismic}. The choice of these attributes depends on the interpretation objective. For instance, several attributes are based on geometric properties \cite{taner1994seismic}, \cite{barnes1992calculation}, \cite{chen1997seismic} while others are derived from the human visual system \cite{shafiq2015detection, shafiq2017salt, shafiq2018role}. More recently, the field has been extended to utilizing deep models as powerful feature extractors \cite{benkert2021explainable, kokilepersaud2022volumetric, wu2019faultseg3d}. Notably, \cite{hyperspectralgrid} propose combining both spectral and spatial features within the feature extractor for hyperspectral image classification.

Within the context of deep learning, our work relates to learning informative representations. Here examples include contrastive feature learning \cite{chen2020simple, he2020momentum} or gradient based representations \cite{spm, prabhushankarintrospective, kwon2020backpropagated}. In addition, recent architectural advancements \cite{vaswani2017attention} gave rise to a set of powerful feature extractors in the form of foundation models. Here, a wide range of foundation models have been explored including multimodal feature learners such as CLIP \cite{clip}, MM1 \cite{mckinzie2024mm1} or LLaVa \cite{lava}; or application specific learners such as segment anything \cite{segmentanything}, dept-anything \cite{depthanything}, and DINOv2 \cite{dinov2}. The trend has since expanded to remote sensing in the form of SpectralGPT \cite{hong2023spectralgpt}. While these works are extract represent powerful feature extractors, they operate on the representation directly which is undesirable when the representation is inaccurate. Our work complements these approaches by extracting information or features from representation shifts rendering a robust alternative for seismic interpretation.


\section{Background}
In this section, we introduce definitions and notations for both active learning and semantic segmentation within the context of seismic interpretation. 
\subsection{Notation and Problem Setup}
In this paper, we view seismic interpretation as a semantic segmentation task. We model a conditional distribution $p^*(\mathbf{y}|\mathbf{x})$ with a deep model $p_w(\mathbf{y}|\mathbf{x}) = h_w(\mathbf{x})$ where $\mathbf{x} \in \mathcal{X} \subset \mathbb{R}^{HxW}$ represents the an in-/crossline from the target volume $\mathcal{X}$. Further, $\mathbf{y} \in \{1, .., K\}^{HxW}$ represents the interpretation label with $K$ subsurface categories and $p_w$ the conditional distribution of the neural network with weight parameters $w$. In practice, the weight parameters are optimized from training data $\mathcal{D} = \{y_i, x_i\}_{i=1}^{N}$ collected from a subset of the full input space $\mathcal{X}_{ID} \subset \mathcal{X}$. Consequently, the target distribution $p^*$ consists of an in-distribution component (ID) $p^*(y|\mathbf{x}, \mathbf{x} \in \mathcal{X}_{ID})$, as well as an out-of-distribution (OOD) $p^*(y|\mathbf{x}, \mathbf{x} \notin \mathcal{X}_{ID})$ component which is not supported by the training data \cite{liu2020simple}:

\begin{equation}
\label{eq:conditional-decomposition}
\begin{split}
   p^*(y|\mathbf{x}) = &p^*(y|\mathbf{x}, \mathbf{x} \in \mathcal{X}_{ID})*p^*(\mathbf{x} \in \mathcal{X}_{ID}) +\\ &p^*(y|\mathbf{x}, \mathbf{x} \notin \mathcal{X}_{ID})*p^*(\mathbf{x} \notin \mathcal{X}_{ID}).
\end{split}
\end{equation}

During training, the model learns the in-distribution component $p^*(\mathbf{y}|\mathbf{x}, \mathbf{x} \in \mathcal{X}_{ID})$ but does not have access to $p^*(\mathbf{y}|\mathbf{x}, \mathbf{x} \notin \mathcal{X}_{ID})$. As a result, predictions on out-of-distribution volumes $\mathbf{x} \notin \mathcal{X}_{ID}$ are arbitrarily incoherent and may significantly differ from the underlying geology.

\subsection{Active Learning}
In active learning \cite{cohn1996active}, the objective involves maximizing neural network performance under minimal data annotations. We consider a dataset $\mathcal{D}$ where the subset $\mathcal{D}_{train}$ represents the current set of annotated sections and $\mathcal{D}_{pool}$ the unlabeled target subset. Using the above notation, the input domain reduces to the union of both $\mathcal{D}_{train}$ and $\mathcal{D}_{pool}$ - i.e. $\mathcal{D}_{train} \cup \mathcal{D}_{pool} = \mathcal{X}_{ID}$ where $\mathcal{D}_{train} \cap \mathcal{D}_{pool} = \emptyset$. In batch active learning, the algorithm selects the most informative batch of $b$ sections $\mathbf{X^*} = \{\mathbf{x^*_1}, ..., \mathbf{x^*_b}\}$ that, when annotated, result in the highest interpretation performance of the network. Formally, we define the selection of $\mathbf{X^*}$ as 

\begin{equation}
\label{eq:aquisition-function}
    X^* = \argmax_{\mathbf{x_1}, ..., \mathbf{x_b} \in D_{pool}} a(\mathbf{x_1}, ..., \mathbf{x_b} | h_w).
\end{equation}

Here, $a$ represents the acquisition function mapping sections individually (or collectively) to a single value of informativeness defined differently across active learning algorithms.




\section{Methodology}
\subsection{Representation Shifts}
\label{sec:representation-shifts}
\begin{figure}[!h]
    \begin{center}
        \includegraphics[scale=0.40]{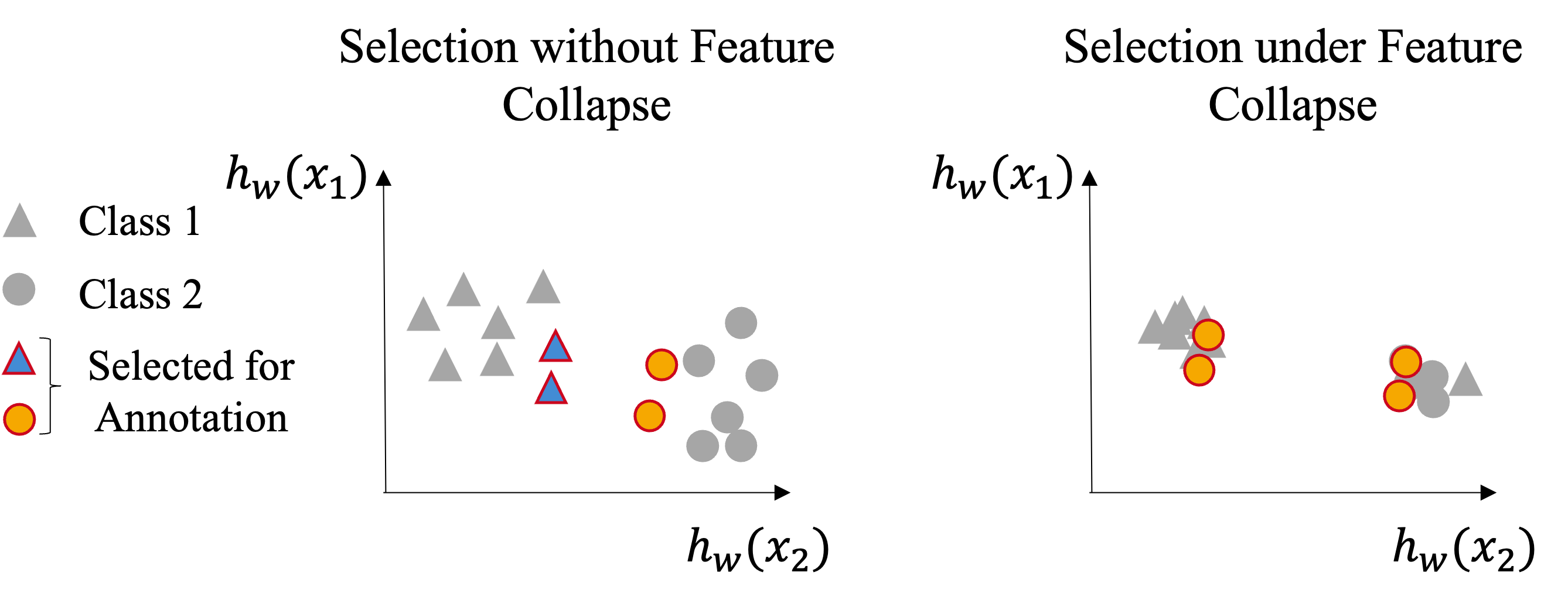}
    \end{center}
    \caption{Toy example of an a acquisition batch selection with an accurate representation (left) and a representation suffering under feature collapse. The collapsed representation results in a significant randomization within the selection algorithm.}
    \label{fig:fcollapse-al-toy}
\end{figure}

Within Equation~\ref{eq:aquisition-function} we observe the direct dependency to the model representation $h_w(\mathbf{x})$. While effective within the general context of computer vision, the characteristic is undesirable for seismic active learning applications:

First, seismic interpretation settings struggle under limited annotated data. 
Due to the high time and monetary cost, sparse training sets with a handful of annotated sections are common settings in seismic interpretation. Within the context of active learning, the characteristic results in significantly less samples within the initial training pool as well as a reduced acquisition batch size. As a result, model representations $h_w(\mathbf{x})$ can overfit to the training domain $\mathcal{D}_{train}$ and collapse knowledge of the domain distribution $p^*(\mathbf{x} \in \mathcal{X}_{ID})$ or the unlabeled target subset $p^*(\mathbf{x} \in \mathcal{D}_{train})$ respectively; A phenomenon referred to as feature collapse \cite{van2020duq}. For machine learning paradigms such as active learning, knowledge of the domain distribution contributes to less randomization within the selection algorithm. We show a toy example of feature collapse in Figure~\ref{fig:fcollapse-al-toy}. On the left, we show representations of a model trained with significant training data. Here, the active learning algorithm accurately selects informative samples (e.g. data points close to the decision boundary). In contrast, when limited data is available the representations can collapse features and results in selection randomization as different points are difficult to distinguish from each other. 

A second major challenge in seismic data is label ambiguity. Specifically, annotations in seismic interpretations can differ depending on the annotator or due to information availability. Consequently, high data noise causes the network to associate structures and features to incorrect facies, rendering the representation inaccurate for active learning. 

In light of the severe disadvantages, a more prudent strategy involves deriving relevant distributional characteristics from the representation indirectly through representation shifts. Within this context, we consider two separate models $h_{w_1}$ and $h_{w_2}$ derived from an optimization process over $D_{train}$. We define a representation shift $\Delta h(\mathbf{x})$ as a suitable distance metric $\mathbf{d}$ between $h_{w_1}(\mathbf{x})$ and $h_{w_2}(\mathbf{x})$:

\begin{equation}
\label{eq:representation-shift}
\begin{split}
   \Delta h(\mathbf{x}) = \mathbf{d}(h_{w_1}(\mathbf{x}), h_{w_2}(\mathbf{x})).
\end{split}
\end{equation}

\begin{figure}
    \begin{center}
        \includegraphics[scale=0.40]{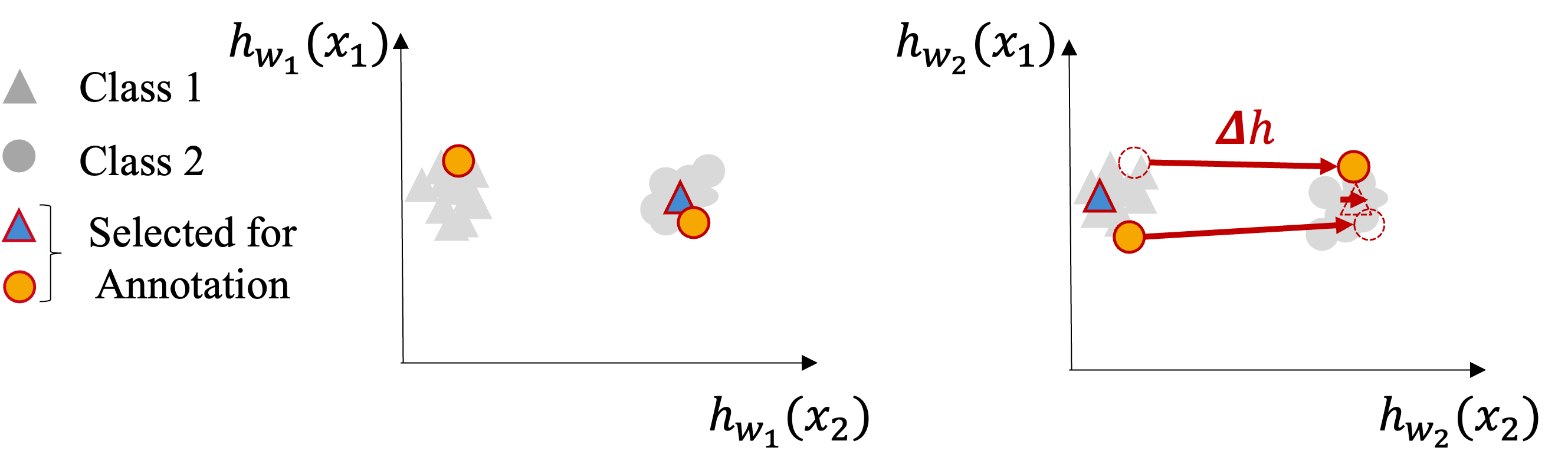}
    \end{center}
    \caption{Toy example of a representation shift under feature collapse applied to a binary classification problem. Left: the model representation of $h_{w_1}(\mathbf{x})$. Right: the model representation of $h_{w_1}(\mathbf{x})$.}
    \label{fig:representation-shift-toy}
\end{figure}

From a machine learning perspective, representation shifts are more robust to feature collapse as they are measured from the model dynamics indirectly instead of originating from a possibly inaccurate data representation $h_w(\mathbf{x})$. To illustrate the robustness process, we show a toy example of representation shifts in Figure~\ref{fig:representation-shift-toy}. While both models $h_{w_1}(\mathbf{x})$ and $h_{w_2}(\mathbf{x})$ are trained on limited data and suffer under feature collapse, we can differentiate data points even though the majority of samples are collapsed to similar positions within the model manifold.

Apart from mathematical advantages, representation shifts have an intuitive interpretation within the context of geophysics. Specifically, we can view $h_{w_1}(\mathbf{x})$ and $h_{w_2}(\mathbf{x})$ as two separate interpretations of the same section and $\mathbf{d}$ as a measure of disagreement between the two. Thus, our approach intuitively models a practical interpretation workflow for geophysical applications.


\subsection{Spatially Aware Active Learning}
\label{sec:spatially-aware-active-learning}

In the previous section, representation shifts were introduced to capture disagreement information in seismic interpretation. Within the following section, representation shifts are applied to active learning to extract regions of geophysical interest from an unlabeled seismic section. Using our previous notation, we reduce the randomization occurring due to feature collapse by focusing our selection on interesting structures or textures within the seismic section. More formally, given a seismic section $\mathbf{x}$, we want to filter relevant geological areas $\phi(\mathbf{x})$ that contain distributional information within the model manifold $h_w(\mathbf{x})$. Applied to Equation~\ref{eq:aquisition-function}, the selection of the next set of sections $X^*$ amounts to

\begin{equation}
\label{eq:aquisition-function-w-filter}
    X^* = \argmax_{\phi(\mathbf{x_1}, h_w), ..., \phi(\mathbf{x_b}, h_w) \in D_{pool}} a(\phi(\mathbf{x_1}, h_w), ..., \phi(\mathbf{x_b}, h_w) | h_w),
\end{equation}

where $\phi: \mathbb{R}^{HxW} \rightarrow \mathbb{R}^{HxW}$ is broadly defined as

\begin{equation}
\label{eq:filter-function}
    \phi(\mathbf{x}, h_w) = \mathbf{x}*m(\mathbf{x}, h_w).
\end{equation}

Here,  $m: \mathbb{R}^{HxW} \rightarrow \{0, 1\}^{HxW}$ represents a binary function that maintains seismic traces containing relevant information and removes redundancies. We refer to paradigms based on Equation~\ref{eq:aquisition-function-w-filter} as \emph{Spatially Aware Active Learning}.
In Figure~\ref{fig:atlas-wf}b, we provide an overview of the spatially aware active learning workflow with the embedded disagreement filter $\phi$. Instead of directly feeding the representation of the entire sectoin into the active learning algorithm, the disagreement filter reduces the active learning input to regions of geophysical interest.

\subsection{Active Transfer Learning for Attention Sensitivity}
\label{sec:atlas-methodology}
\begin{figure*}[!t]
    \begin{center}
        \includegraphics[scale=0.33]{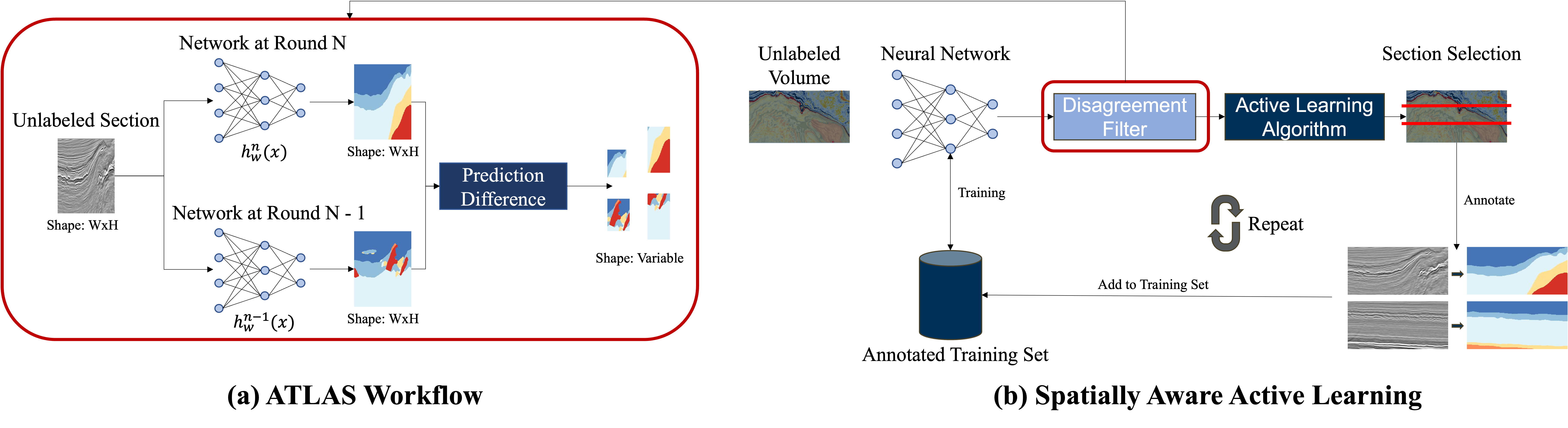}
    \end{center}
    \caption{\textbf{a)} high-level workflow of our plug-in active learning framework, \texttt{ATLAS}. For each unlabeled section, we extract the predictions from both the current round as well as the previous round and derive the prediction difference. Subsequently, we filter regions where the predictions conflict, and process the disagreeing regions by the active learning algorithm exclusively. \textbf{b)} spatially-aware active learning workflow. The active learning framework is modified by introducing a disagreement filter that restricts the input of the active learning algorithm to regions of geological interest.}
    \label{fig:atlas-wf}
\end{figure*}

Within this subsection, we propose a concrete implementation of the filter $\phi$ with representation shifts. We illustrate the workflow in Figure~\ref{fig:atlas-wf}a. For every unlabeled section, we extract the model predictions from the current, and previous active learning round ($n$ and $n-1$ respectively) and perform a prediction comparison. We filter regions where the interpretations disagree and allow the acquisition function to process sections based on the conflicting regions exclusively. We consider our algorithm an added attention mechanism to regions of geophisical interest and call our plug-in method \emph{Active Transfer Learning for Attention Sensitivity} or \texttt{ATLAS} in short. 
Within the following paragraphs, we formalize the concept with an explicit expression for $\phi$.

Within the context of Equation~\ref{eq:representation-shift}, deriving the prediction difference is equivalent to monitoring representation shifts with respect to individual traces or pixels. Applied to the entire section, $\Delta h(\mathbf{x}) \in \{0, 1\}^{HxW}$ reduces to a binary heatmap with each pixel $k$ indicating a prediction switch or conflict in between the predictions of both active learning rounds. More formally, we define an individual trace $k \in \{0, ..., H*W - 1\}$ of $\Delta h$ as

\begin{equation}
\label{eq:switch-events}
\Delta h(\mathbf{x})_k = int(\tilde{y}^n_k(\mathbf{x}) \neq \tilde{y}^{n-1}_k(\mathbf{x})) = s_k(\mathbf{x}).
\end{equation}

Here, $\tilde{y}^n_k(\mathbf{x}) = \argmax_{c \in \{1, ..., K\}} h_{w_n}^c(\mathbf{x})$ represents the predicted class when using the fully trained model from the $n$-th active learning round $h_{w_n}$.  Consequently, we define $\phi$ as a multiplication with the binary heatmap $\Delta h(\mathbf{x})_k$:




\begin{equation}
\label{eq:phi-atlas}
    \phi(\mathbf{x}, h_w) = \mathbf{x}*\Delta h(\mathbf{x}).
\end{equation}

We emphasize that Expression~\ref{eq:representation-shift} represents a generalized version of the explicit implementation in \texttt{ATLAS}. Specifically, $h_{w_1}$ and $h_{w_2}$ translates to $h_{w_n}$ and $h_{w_{n-1}}$ respectively and the distance metric $\mathbf{d}$ is represented by the binary prediction switch image $\mathbf{s}$. While beyond the scope of this paper, we note that different implementations of $\Delta h$ and $\phi$ are possible and left for future research.

\section{Experiments}
\label{sec:results}
Within this section, we address the following questions: 1) how does \texttt{ATLAS} perform in combination with several active learning acquisition functions? 2) are improvements consistent across several test volumes?; and 3) which regions does \texttt{ATLAS} generally target?

\subsection{Experimental Setup}

\begin{figure}[!h]
    \begin{center}
        \includegraphics[scale=0.37]{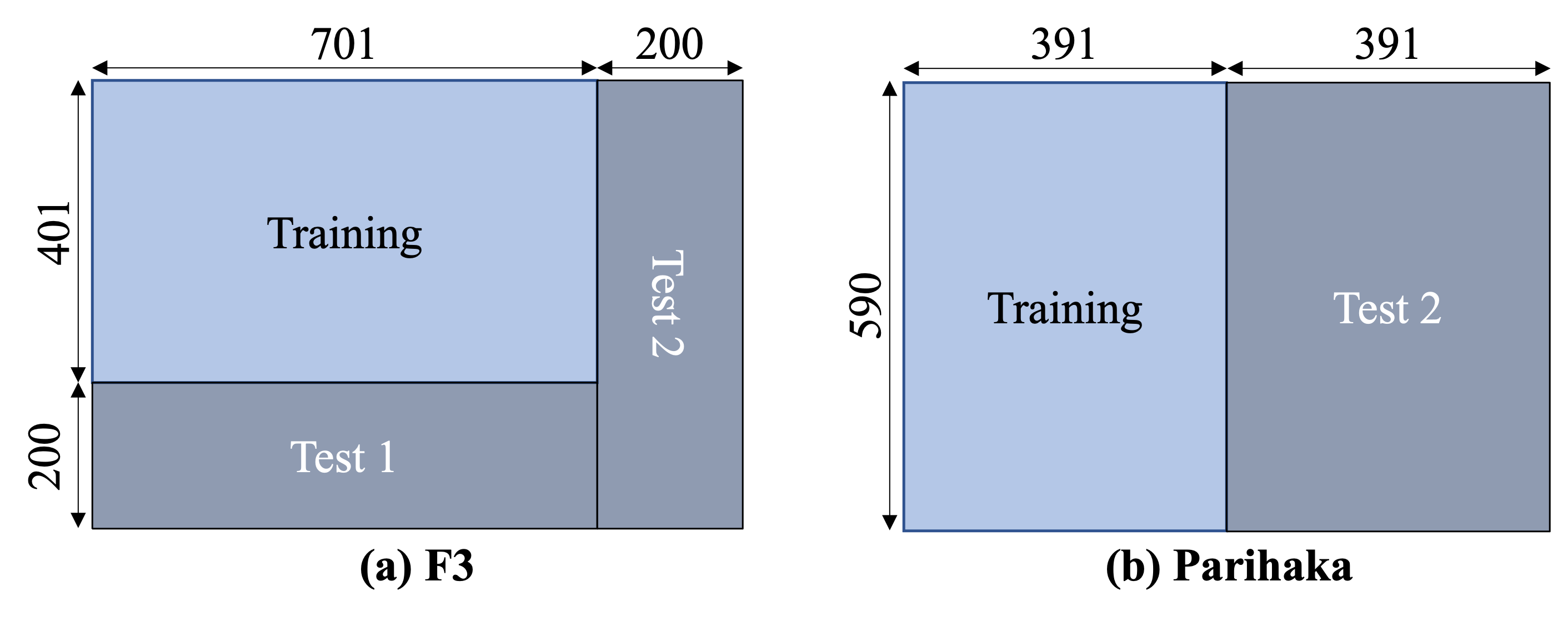}
    \end{center}
    \caption{Training and test split of both seismic datasets. \textbf{a)} The F3 dataset located in the Netherlands. We use the split of \cite{alaudah2019machine} with one contiguous training, as well as two test volumes. \textbf{b)} The Parihaka volume from New Zealand. We split the volume into one contiguous training and test volume.}
    \label{fig:training-test-split}
\end{figure}

For our active learning experiments, we select two random sections within the training volume as our initial $D_{train}$ and select two additional sections each active learning round for annotation. We select a small acquisition batch size of two as it represents a realistic magnitude for the application of seismic interpretation. For our active learning benchmark, we analyze the performance of \texttt{ATLAS} by comparing different acquisition functions with and without applying \texttt{ATLAS}. Specifically, we compare \texttt{ATLAS} to ranking informativeness based on the entire section $\mathbf{x}$ instead of the spatially targeted subsection $\phi(\mathbf{x})$. In all experiments, our architecture is based on the deeplab-v3 architecture first proposed by \cite{chen2017deeplab}. If not otherwise specified, we implement the backbone with a resnet-18 \cite{he2015deep} feature extractor, and optimize using the adam variant of stochastic gradient descent with a learning rate of $5e^{-5}$. During each active learning round, we train the model until it achieves an at least an mean-intersection-over-union (mIoU) performance of $0.9$ and log performance on the test volume in overall mIoU, as well as class accuracy. All experiments are conducted over five separate random seeds.

In this paper, experiments are provided on two different seismic data volumes. The first is the F3 benchmark located in the Netherlands \cite{alaudah2019machine}, with annotations corresponding to six different facies categories. To facilitate conditions of the original paper, we use the same experimental setup as the original paper: we use the \hyperlink{https://github.com/olivesgatech/facies_classification_benchmark}{pre-processed volume} provided by the original implementation, and split the volume according to the initial benchmark. The partition contains one contiguous training set, and two test volumes (Figure~\ref{fig:training-test-split}a). The second volume is the publicly available Parihaka dataset located in New Zealand. The seismic data is provided by New Zealand Petroleum and Minerals, while the associated labels are provided from Chevron U.S.A. Inc., and are licensed under \hyperlink{https://creativecommons.org/licenses/by-sa/4.0}{CC BY-SA 4.0}. Similar to F3, the labels include annotations for six different facies. Before training, the seismic dataset is clipped within the range of $[-1500, 1500]$, and sub-sampled in depth direction by factor three. Further, we remove the first 200 and deepest 42 depth slices as they contain the same classes throughout the volume and are not informative to the training process. As annotations for the original test volume are not publicly available, we split the original training volume as shown in Figure~\ref{fig:training-test-split}b. We choose this partition due to class representation and to facilitate structural variability between training and test volume, making the dataset more challenging. We make our code publicly available at PLACEHOLDER.

\begin{table*}[!t]
	\centering
	\caption{Average accuracy, and standard deviation on the F3 dataset. }
    \vskip 0.15in
	\begin{tabular}{ |p{2.3cm}||p{1.6cm}|p{1.6cm}|p{1.6cm}|p{1.6cm}|p{1.6cm}|p{1.6cm}|p{1.6cm}|}
		\hline
		
		\multicolumn{1}{|l||}{Algorithms} & \multicolumn{1}{|c|}{\centering mIoU}& \multicolumn{1}{|c|}{\centering Upper N.S.}& \multicolumn{1}{|c|}{\centering Mid. N.S.}& \multicolumn{1}{|c|}{\centering Lower N.S.}&  \multicolumn{1}{|c|}{\centering Chalk.}& \multicolumn{1}{|c|}{\centering Scruff} & \multicolumn{1}{|c|}{\centering Zechstein}\\
        \hline
    	Entropy           & 0.591 ± 0.014      & \bf{0.975 ± 0.001} & 0.846 ± 0.012      & 0.950 ± 0.003      & 0.636 ± 0.009      & 0.373 ± 0.042      & \bf{0.461 ± 0.010}             \\
		ATLAS Entropy     & \bf{0.620 ± 0.009} & \bf{0.976 ± 0.001} & \bf{0.872 ± 0.007} & \bf{0.957 ± 0.001} & \bf{0.651 ± 0.014} & \bf{0.495 ± 0.028} & 0.417 ± 0.071      \\
		\hline
    	Margin            & 0.574 ± 0.017      & \bf{0.977 ± 0.002} & 0.840 ± 0.012      & 0.951 ± 0.003      & 0.643 ± 0.012      & 0.305 ± 0.079      & 0.424 ± 0.055             \\
		ATLAS Margin      & \bf{0.612 ± 0.005} & \bf{0.976 ± 0.001} & \bf{0.858 ± 0.013} & \bf{0.952 ± 0.002} & \bf{0.652 ± 0.016} & \bf{0.440 ± 0.019} & \bf{0.500 ± 0.025}      \\
		\hline
    	Least Conf.       & 0.575 ± 0.022      & \bf{0.970 ± 0.005} & 0.808 ± 0.025      & 0.944 ± 0.010      & 0.637 ± 0.022      & 0.388 ± 0.041      & 0.430 ± 0.046             \\
		ATLAS Least Conf. & \bf{0.619 ± 0.006} & \bf{0.974 ± 0.003} & \bf{0.869 ± 0.006} & \bf{0.956 ± 0.002} & \bf{0.653 ± 0.013} & \bf{0.480 ± 0.026} & \bf{0.459 ± 0.046}      \\
		\hline
	\end{tabular}
	\label{table:results-aa}
\end{table*}

\subsection{Numerical Experiments}

In active learning, the sampling performance can vary across different rounds. 
For this purpose, we summarize the overall performance of an active learning algorithm by using the average accuracy metric \cite{benkert2022forgetful, benkert2023gaussian}. Specifically, we integrate over the accuracy curve and normalize by the number of acquired sections. Specifically, given a total number of rounds $R$, as well as the evaluation metric $metric(r)$ for every round $r$, the average accuracy $P_{acc}$ is defined as

\begin{equation}
\label{eq:ac}
    P_{acc} = \frac{1}{R} \int_r metric(r) \,dr.
\end{equation}

For our numerical analysis, we benchmark three popular active learning acquisition functions in combination with \texttt{ATLAS}: entropy sampling \cite{wang2014new}, margin sampling \cite{roth2006margin}, and least confidence sampling \cite{wang2014new}. We choose these strategies as they represent popular difficulty based active learning approaches that are of particular interest to seismic interpretation as areas of geological interest are frequently concentrated in regions if high learning difficulty \cite{benkert2022example}. In Table~\ref{table:results-aa} and Table~\ref{table:results-aa-parihaka}, we report our results on average accuracy in mIoU, as well as class-wise accuracy across the F3 block, and Parihaka dataset respectively. 
We show the average accuracy for each metric in mean and standard deviation. 

In all settings, \texttt{ATLAS} results in either an improvement or a statistically insignificant difference to the baseline. The observation reflects in the overall mIoU performance across all classes. 
For each acquisition function, \texttt{ATLAS} shows a significant improvement of up to approximately 4\% on the F3 block, and 12\% on the Parihaka volume. The improvements are agnostic to class frequency in the volume: 
for F3, class accuracy is increased both within common classes such as the middle and lower north sea group, as well as underrepresented such as scruff; the rarest class within both training and test sets. 
We reason that regions with high representation shifts are particular common around class boundaries, regardless of class frequency. Therefore, \texttt{ATLAS} targets structurally interesting regions in both well represented, as well as less represented classes. For the Parihaka volume, \texttt{ATLAS} boosts performance in nearly every class by a significant margin; even more so as on the F3 dataset. We attribute the strong performance to the structure of the facies. Within the F3 block, rare classes are concentrated in salt domes or are densely located in deeper regions with less surface boundaries between other classes. In contrast, the Parihaka volume contains underrepresented facies (e.g. slope valley) that are thin layers between well represented classes. As \texttt{ATLAS} emphasizes class boundaries it is more effective in selecting geologically different regions on the Parihaka dataset due to the large surface boundaries of underrepresented classes. We perform a detailed analysis of the learning curves in Section~\ref{sec:learning-curves}.


\subsection{Learning Curves}
\label{sec:learning-curves}
\begin{figure*}[!t]
    \begin{center}
        \includegraphics[scale=0.37]{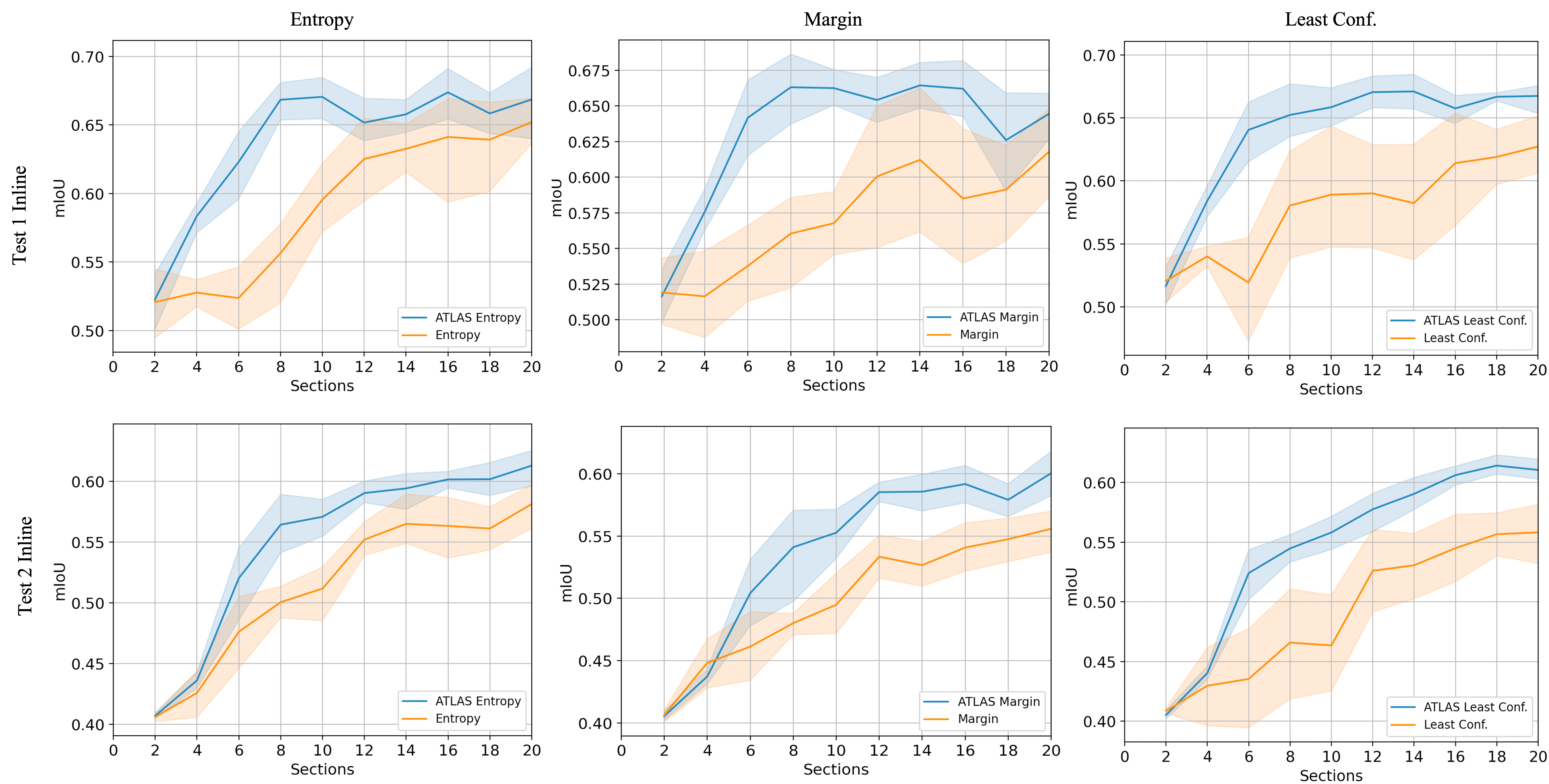}
    \end{center}
    \caption{Learning curves of entropy, margin, and least confidence sampling in combination with \texttt{ATLAS} on the F3 dataset. We show the overall mIoU performance on test set one in the first row, and test set two in the second.}
    \label{fig:lc-strat-comp}
\end{figure*}

\begin{figure*}
    \begin{center}
        \includegraphics[scale=0.37]{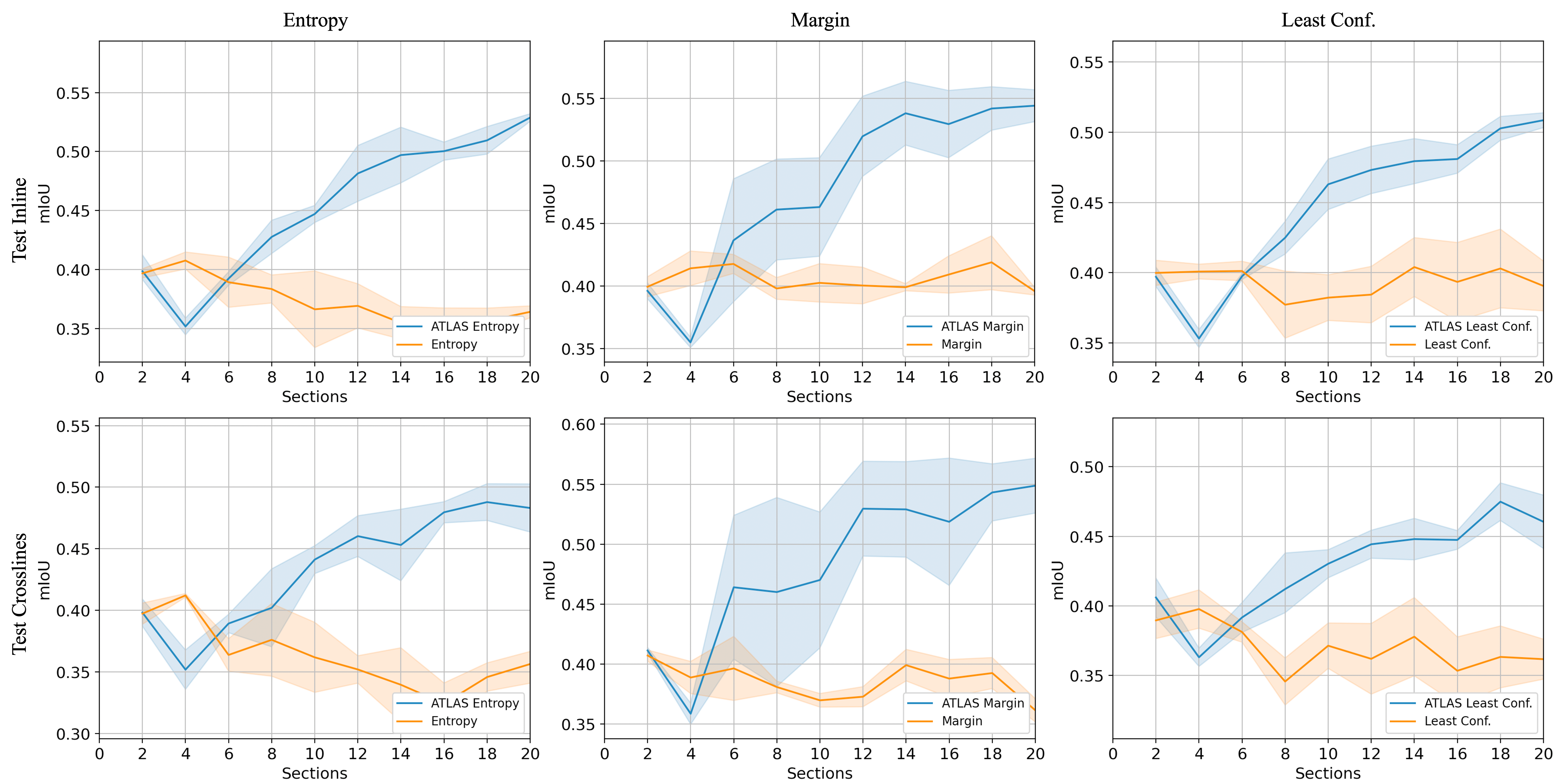}
    \end{center}
    \caption{Learning curves of entropy, margin, and least confidence sampling in combination with \texttt{ATLAS} on the Parihaka dataset. We show the overall mIoU performance on test set inlines in the first row, and test set crosslines in the second.}
    \label{fig:lc-strat-comp-parihaka}
\end{figure*}

Within this subsection, we analyze the explicit learning curves of the individual active learning algorithms in combination with \texttt{ATLAS}. In Figure~\ref{fig:lc-strat-comp}, we show the total mIoU on the inlines of both the first and second test volume of the F3 block, where the second represents the more difficult of the two. Figure~\ref{fig:lc-strat-comp-parihaka}, shows the mIoU on the Parihaka test volume inlines and crosslines. The rows present test set one and two, or in-/crossline respectively, while the columns show the individual algorithms entropy, margin, and least confidence with and without \texttt{ATLAS}. The shaded region represents the 95\% confidence interval with respect to the five random seeds.

Our plots support our numerical analysis. In all experiments, \texttt{ATLAS} matches or improves the baseline performance in all three algorithms in nearly every round. Within the F3 volume, \texttt{ATLAS} is particularly effective on the simpler test set in early rounds, and on the difficult test 2 volume in later rounds. For instance, deploying \texttt{ATLAS} on the margin algorithm results in a significant improvement on test set one in rounds four to eleven, while \texttt{ATLAS} strongly outperforms on test two in round 8 and onwards. We reason that switches occur frequently in simple structures within early active learning rounds as complex structures are more challenging with small training sets. As a result, common structures result in significant shifts and are queried by \texttt{ATLAS} rendering a strong performance on the less difficult test one volume. In contrast, rare structures or classes are not sufficiently represented enough in early rounds and the model consistently misclassifies difficult samples \cite{benkert2022example}. Therefore, difficult structures are queried in later rounds when the model representation is mature with more representation shifts. On the Parihaka dataset, the baseline algorithm and \texttt{ATLAS} follow different trends. \texttt{ATLAS} increases the mIoU, while the baseline strategy stagnates or even reduces the performance. We attribute the behavior to the structure of the facies in the Parihaka dataset. \texttt{ATLAS} is effective on volumes where geophysically interesting regions share large class boundaries. For facies that exhibit a thin layer structure as is present in Parihaka, representation shifts concentrate around thin underrepresented classes giving the regions more weight in the selection process. In contrast, the baseline algorithm evaluates the acquisition function on the entire section, biasing the selection process to well-represented regions. Note that both strategies eventually converge to the same performance when the entire training volume is queried. In Section~\ref{sec:results-predictions}, we provide an additional analysis of \texttt{ATLAS} by considering the network predictions. 

\begin{figure*}[!t]
    \begin{center}
        \includegraphics[scale=0.35]{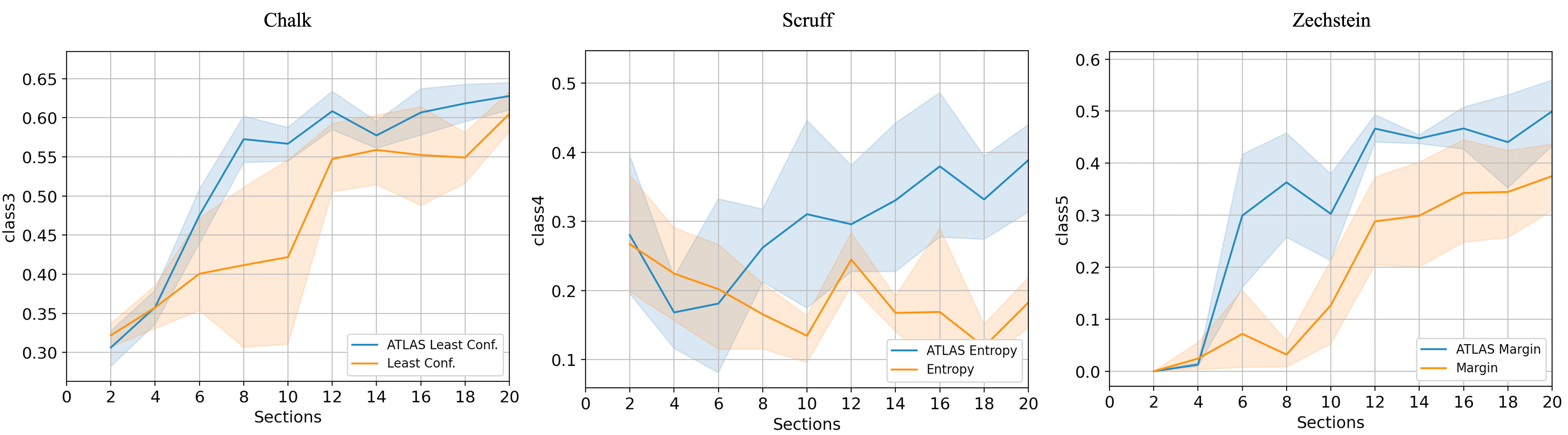}
    \end{center}
    \caption{Class accuracy learning curves of entropy, least confidence, and margin sampling in combination with \texttt{ATLAS}. We show the class accuracy of the underrepresented classes chalk, scruff and zechstein.}
    \label{fig:lc-class-acc}
\end{figure*}

In addition to overall mIoU, we report the class accuracy learning curves of chalk, scruff, and zechstein classes from the F3 block in Figure~\ref{fig:lc-class-acc}. For simplicity, we report chalk using least confidence, scruff with entropy, and zechstein with margin sampling even though the trend is largely consistent accross the volume. In Figure~\ref{fig:lc-class-acc}, we show test two inlines exclusively as it represents the more difficult challenge with complex structures. Similar to our previous observations, our results are consistent with both the learning curves on overall mIoU, as well as our numerical experiments in Table~\ref{table:results-aa}. \texttt{ATLAS} either matches or improves the baseline algorithm even in underrepresented settings.

\begin{table*}[!t]
	\centering
	\caption{Average accuracy, and standard deviation on the Parihaka dataset. }
    \vskip 0.15in
	\begin{tabular}{ |p{2.3cm}||p{1.6cm}|p{1.6cm}|p{1.6cm}|p{1.6cm}|p{1.6cm}|p{1.6cm}|p{1.6cm}|}
		\hline
		
		\multicolumn{1}{|l||}{Algorithms} & \multicolumn{1}{|c|}{\centering mIoU}& \multicolumn{1}{|c|}{\centering Basement}& \multicolumn{1}{|c|}{\centering Mudstone A}& \multicolumn{1}{|c|}{\centering Mass Tr. Dep.}&  \multicolumn{1}{|c|}{\centering Mudstone B}& \multicolumn{1}{|c|}{\centering Slope Valley} & \multicolumn{1}{|c|}{\centering Canyon Sys.}\\
        \hline
    	Entropy           & 0.364 ± 0.013      & 0.932 ± 0.011 & 0.752 ± 0.017      & 0.335 ± 0.071      & 0.620 ± 0.011      & 0.133 ± 0.007      & \bf{0.256 ± 0.123}             \\
		ATLAS Entropy     & \bf{0.481 ± 0.003} & \bf{0.967 ± 0.003} & \bf{0.797 ± 0.002} & \bf{0.644 ± 0.023} & \bf{0.738 ± 0.014} & \bf{0.266 ± 0.022} & \bf{0.387 ± 0.011}      \\
		\hline
    	Margin            & 0.400 ± 0.014      & 0.919 ± 0.029 & 0.777 ± 0.009      & 0.305 ± 0.081      & 0.720 ± 0.008      & 0.112 ± 0.080      & 0.370 ± 0.098             \\
		ATLAS Margin      & \bf{0.518 ± 0.028} & \bf{0.969 ± 0.002} & \bf{0.808 ± 0.008} & \bf{0.687 ± 0.053} & \bf{0.775 ± 0.013} & \bf{0.374 ± 0.064} & \bf{0.422 ± 0.049}      \\
		\hline
    	Least Conf.       & 0.381 ± 0.022      & 0.895 ± 0.029 & 0.762 ± 0.004      & 0.290 ± 0.042      & 0.689 ± 0.033      & 0.134 ± 0.032      & 0.292 ± 0.050             \\
		ATLAS Least Conf. & \bf{0.478 ± 0.006} & \bf{0.969 ± 0.003} & \bf{0.796 ± 0.005} & \bf{0.653 ± 0.021} & \bf{0.735 ± 0.007} & \bf{0.269 ± 0.016} & \bf{0.360 ± 0.010}      \\
		\hline
	\end{tabular}
	\label{table:results-aa-parihaka}
\end{table*}

\subsection{Predictions}
\label{sec:results-predictions}
\begin{figure*}
    \begin{center}
        \includegraphics[scale=0.30]{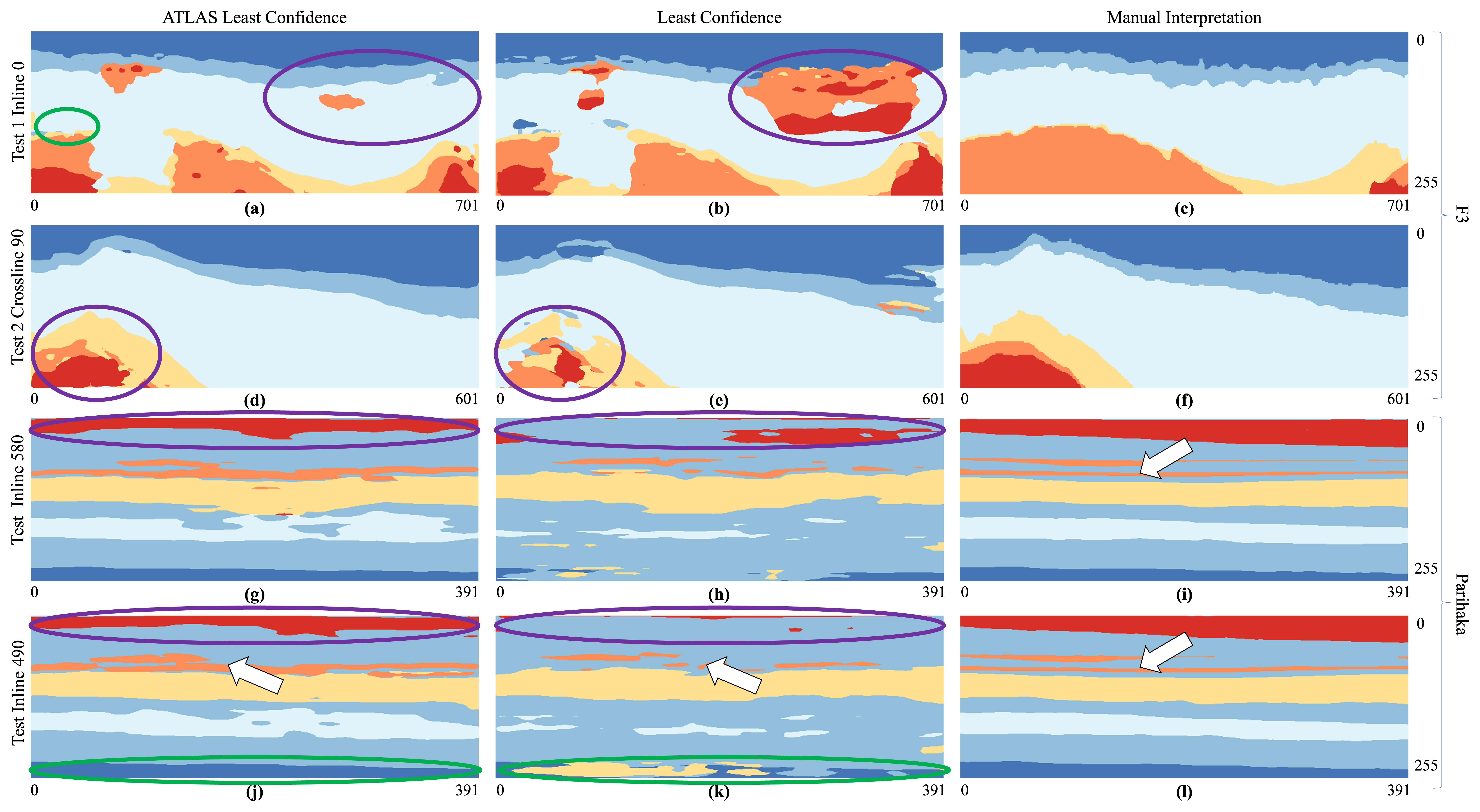}
    \end{center}
    \caption{Network predictions in comparison to manual interpretation for the F3 block and Parihaka dataset. The first two rows show predictions in round 14 from the F3 dataset, the bottom two rows are predictions from round seven on the Parihaka dataset. We compare least confidence sampling to least confidence in combination with \texttt{ATLAS}. The first column contains least confidence in combination with \texttt{ATLAS}, the second contains the least confidence baseline, and the third the manual interpretation by an expert.}
    \label{fig:predictions-lconf}
\end{figure*}

In addition to numerical metrics, we directly compare the network predictions of \texttt{ATLAS} with the baseline algorithm. In Figure~\ref{fig:predictions-lconf}, we show the network predictions of least confidence sampling, as well as least confidence in combination with \texttt{ATLAS}. 
To facilitate accurate comparison, the manual interpretation is plotted in the third column. From a machine learning perspective, manual interpretation represents the target that the model aims to approximate with maximum precision. All predictions depicted in Figure~\ref{fig:predictions-lconf} are derived from round seven for the Parihaka dataset, and round 14 for the F3 block.
Overall, we note an improved segmentation performance within the output predictions. The baseline algorithm exhibits significantly more semantically incorrect predictions and fails in complicated structures. In contrast, adding \texttt{ATLAS} reduces mispredictions and results in less sporadic outputs. For the F3 block, the baseline algorithm on inline zero in test set one predicts a large scruff (orange class) and zechstein (red class) region in the top right corner even though both classes exclusively appear in deeper regions at the bottom of the section (Figure~\ref{fig:predictions-lconf}b violet circle). In combination with \texttt{ATLAS}, the misprediciton nearly disappears (Figure~\ref{fig:predictions-lconf}a violet circle). Further we note that the horizon predictions in well represented classes such as the upper, middle, and lower north sea groups are spatially consistent and less sporadic. This is particularly visible in Figure~\ref{fig:predictions-lconf}a-b, over the left large scruff mound (green circle). The baseline algorithm predicts the upper north sea group (dark blue) directly next to chalk and above the scruff mound; a incorrect prediction as the middle north sea group occurs in between the upper and lower north sea group exclusively. 
The mispredictions are resolved when utilizing \texttt{ATLAS}. The analysis of the crossline from test volume two (Figure~\ref{fig:predictions-lconf}d-e) reveals that networks trained with \texttt{ATLAS} exhibit superior performance in accurately representing intricate structures. The salt dome structure located on the bottom left (violet circle) is depicted with greater accuracy when trained using \texttt{ATLAS}, as opposed to the baseline approach. On the Parihaka dataset, we note the clear difference in facies structure compared to the F3 block. Rare classes such as slope valley are located in shallower regions and represent layers instead of dense mounds (Marked by the white arrow in Figure~\ref{fig:predictions-lconf}j, and l). As a result the class boundary areas are larger than F3 resulting in stronger representation shifts in geologically interesting regions. In the predictions the effect manifests in significantly improved performance. In Figure~\ref{fig:predictions-lconf}g, h, j, and k the structure of the canyon system (red; marked by violet circle) and basement (dark blue; marked by green circle) is thoroughly visible and less sporadic with \texttt{ATLAS}. Further, underrepresented classes such as slope valley (orange; Figure~\ref{fig:predictions-lconf}j, and k white arrow) are significantly better preserved with \texttt{ATLAS} indicating a more effective sampling framework.

\subsection{Representation Shifts}

\begin{figure*}
    \begin{center}
        \includegraphics[scale=0.45]{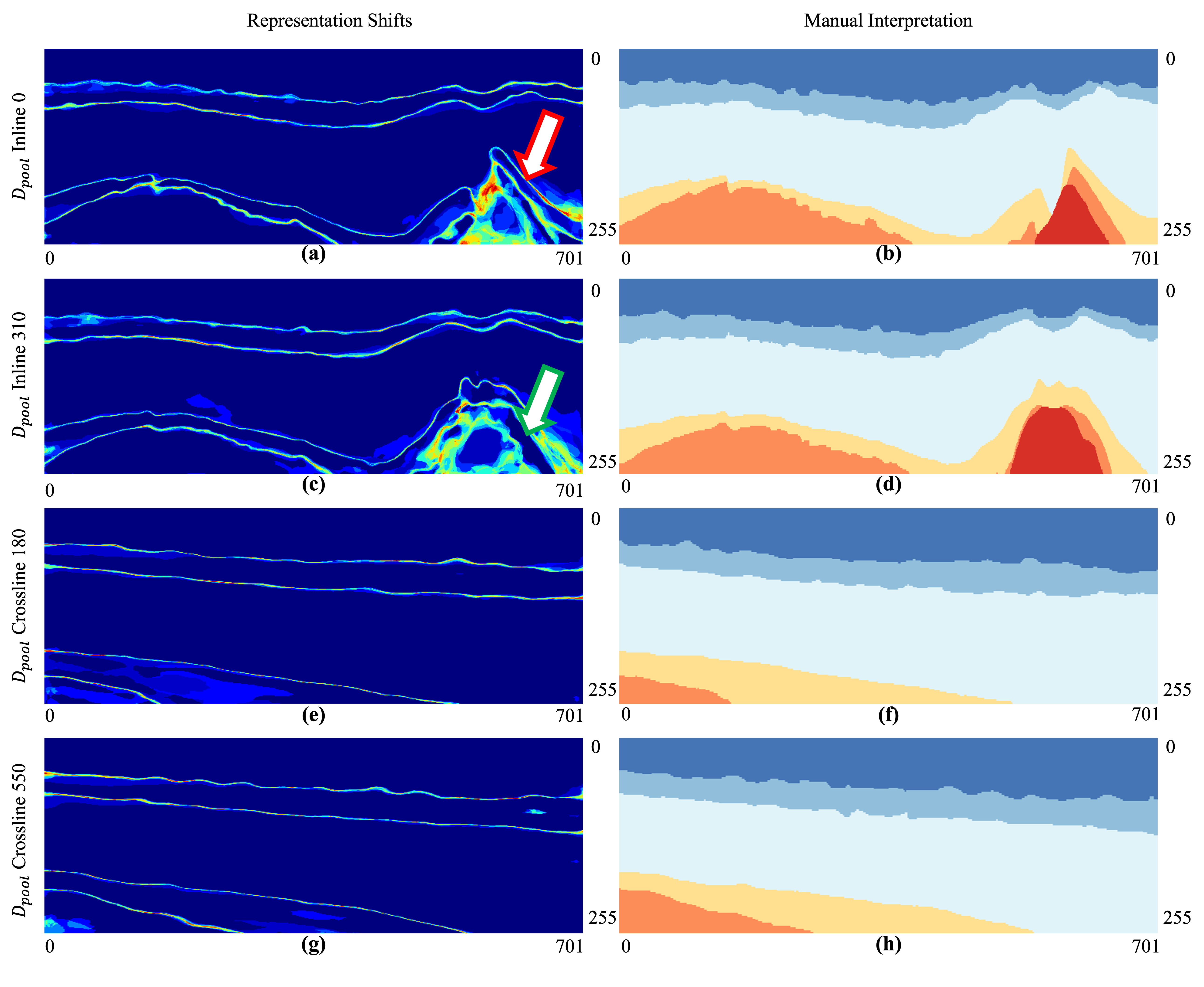}
    \end{center}
    \caption{Number of representation shifts measured in prediction switches per pixel. Red regions represent large amounts of prediction switches, while blue regions represent the opposite. The first two rows show the unlabeled volume $D_{pool}$ in inline zero and 310 respectively, while row three and four show crossline 180 and 550. The first column shows the representation shift heatmpas in round 20 as produced by least confidence sampling with \texttt{ATLAS}, the second depicts the manual interpretation label. We further highlight regions of high representation shift density with a white arrow. \texttt{ATLAS} focuses on these regions within the selection, while conventional active learning averages over the entire section.}
    \label{fig:switches-lconf}
\end{figure*}

Within this subsection, we visualize the regions \texttt{ATLAS} selects. For this purpose, we monitor the prediction switches for the unlabeled pool and accumulate the shifts of each pixel across the entire active learning experiment. 
In Figure~\ref{fig:switches-lconf}, the heatmaps represent the accumulation of prediction switches over 21 active learning rounds with least confidence sampling. All heatmaps originate from the F3 block. We show both sections with interesting structures and underrepresented classes, as well as simpler sections that are easier for the network to learn. Within the Figure, large amounts of representation shifts are color-coded as red while blue indicates the opposite. We note, that prediction switches largely occur either in class boundaries or difficult structures and underrepresented classes. For instance, in Figure~\ref{fig:switches-lconf}a, and c, the salt dome his highlighted, sub-regions with complex structures: in Figure~\ref{fig:switches-lconf}a, the boundary in between zechstein and the middle north sea group is brightly highlighted (white arrow with red contours). A similar observation can be made for Figure~\ref{fig:switches-lconf}c, where the salt dome is highlighted (white arrow with green contours). \texttt{ATLAS} focuses on these areas, while the baseline algorithms average over the entire section. The highlighted regions further support our performance analysis and provide a conceptual explanation to the strong performance of \texttt{ATLAS}.

\section{Summary and Discussion}

\subsection{Summary}


In this paper, we propose incorporating interpretation disagreement in data selection pipelines for deep models in seismic interpretation. 
First, the paper establishes a definition of information content for manual interpretation workflows in the form of disagreement among interpreters. The definition is grounded in a machine learning context by considering the shift between mathematical representations as disagreement among deep models. Combined with active learning, the paper develops a novel framework to select the most informative sections of an unlabeled seismic volume to maximize the generalization performance. Finally, a concrete implementation of the framework is proposed called \texttt{ATLAS}. 
In conclusion, \texttt{ATLAS} has led to enhancements in the data selection process for seismic interpretation. 
We conducted comprehensive experiments to evaluate the performance of \texttt{ATLAS} and our findings indicate a noteworthy enhancement of up to 10\% when compared to the baseline active learning strategy.

\subsection{Discussion and Future Research}
A key observation we made in this paper is that utilizing representation shifts rather than direct network outputs can benefit the generalizability of neural networks in seismic interpretation. We investigated one concrete form of representation shift, namely disagreement to identify relevant regions within a seismic section. While we benchmark our method within the context of data selection, we emphasize that our method can be generalized to a wide range of applications within seismic interpretation. For instance, disagreement may provide an effective measure of uncertainty or of performance degradation. Further, we note that our investigation only focused on one form of representation shifts, namely disagreement with prediction switches. While our initial results are encouraging, we note that our implementation reduces representation shifts to a binary prediction switch which results in a significant information loss. In this context, we see another important research direction in investigating other forms of representation shifts that are non-binary and could provide a richer understanding of the deep neural network.
\bibliographystyle{IEEEtran}

\bibliography{mybib}

\newpage
\section{Acknowledgements}
This work is supported by the Industry Supported ML4Seismic Consortium at Georgia Tech.

\section{Biography Section}
 
\vspace{11pt}

\begin{IEEEbiography}[{\includegraphics[width=1in,height=1.25in,clip,keepaspectratio]{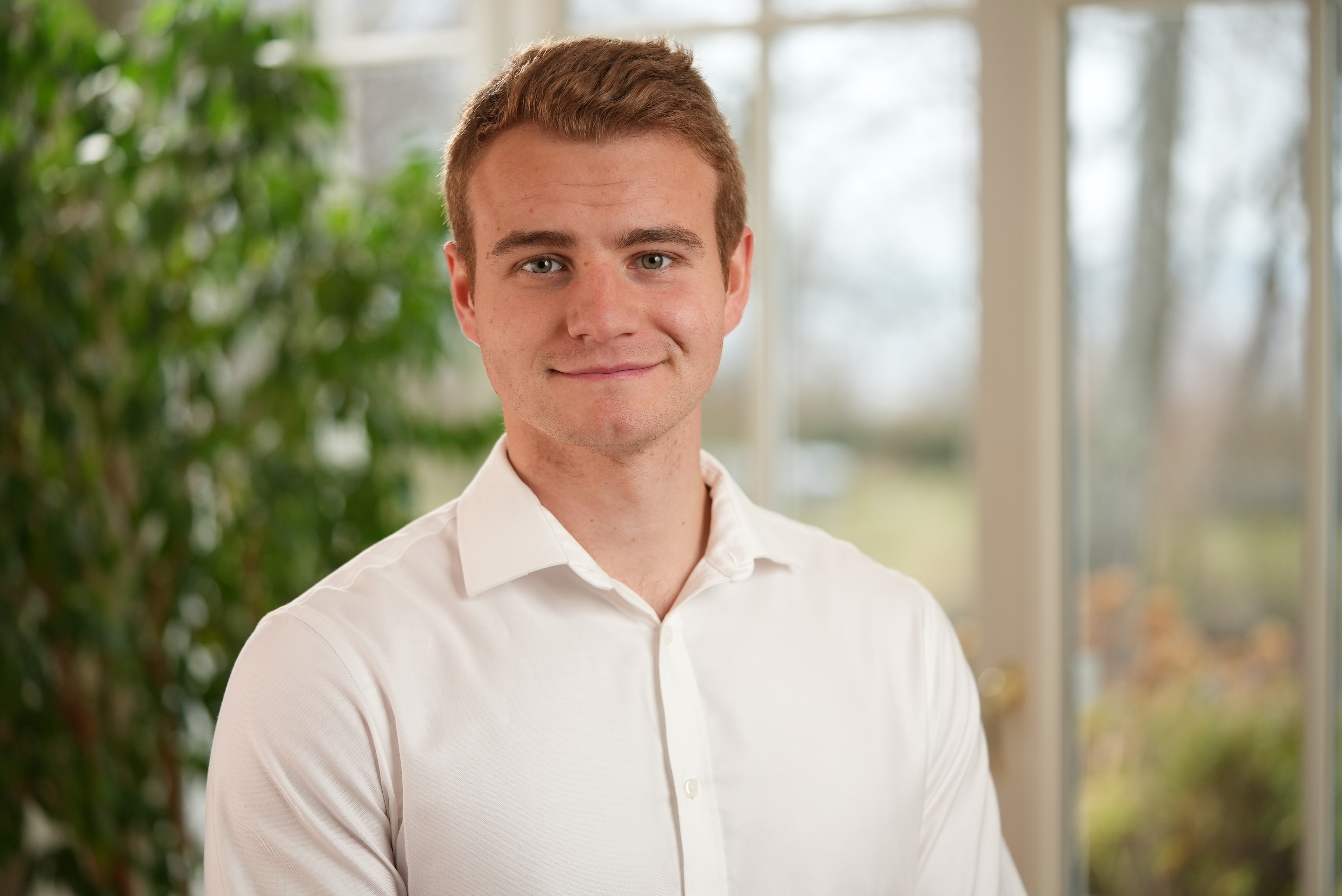}}]{Ryan Benkert}
received his Ph.D. degree in electrical engineering from the Georgia Institute of Technology (Georgia Tech), Atlanta, Georgia, 30332, USA, in 2024. He is currently a deep learning software engineer at NVIDIA in Santa Clara, CA. His research interests are at the intersection of active learning, uncertainty estimation, and performance consistency in neural network learning. Prior to Georgia Tech, he received his B.Sc and M.Sc from the RWTH Aachen University in Germany. He is the recipient of the Otto-Junker award for his achievements during his graduate studies at the RWTH Aachen in Germany. 

%
\end{IEEEbiography}

\begin{IEEEbiography}[{\includegraphics[width=1in,height=1.25in,clip,keepaspectratio]{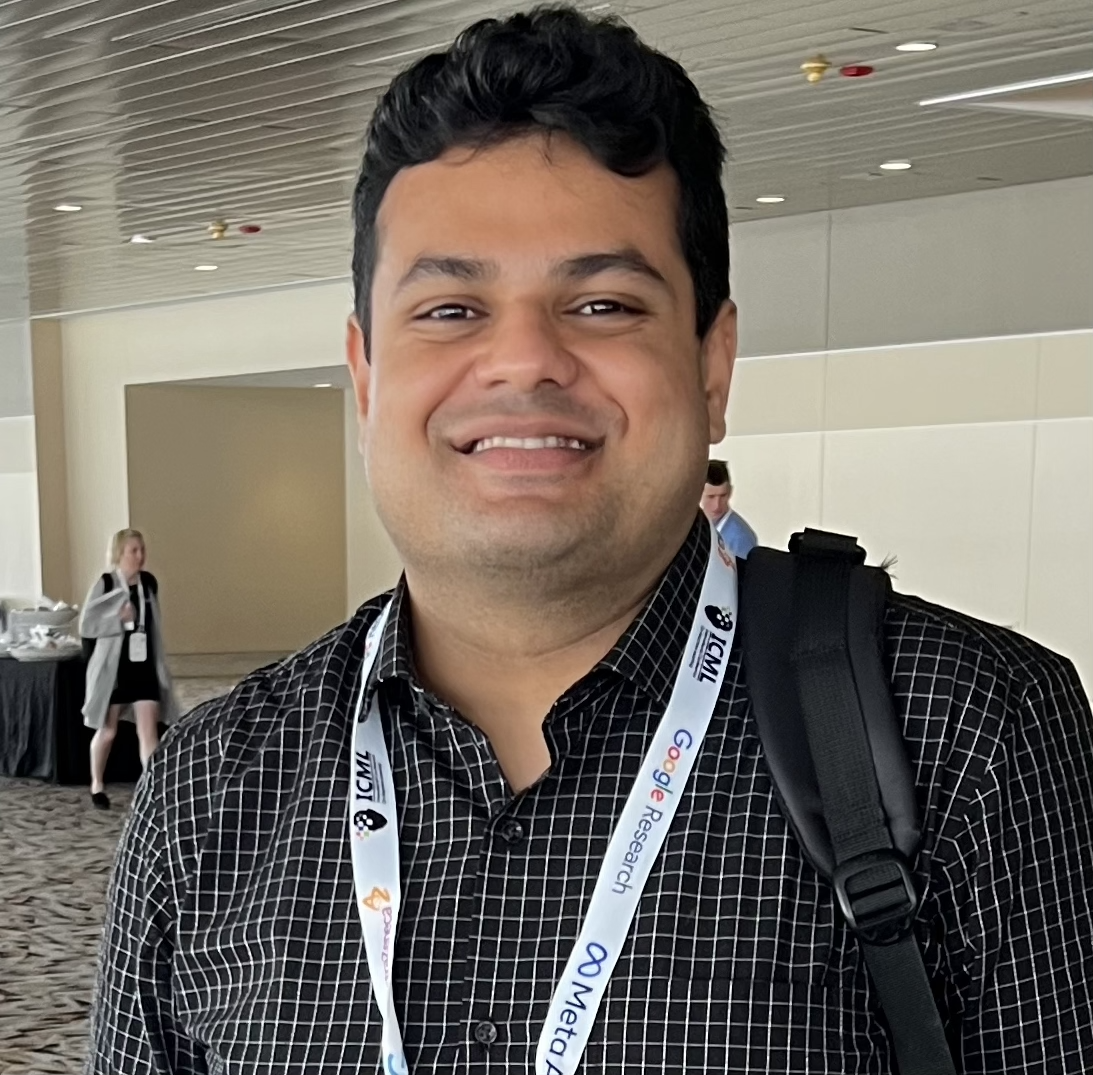}}]{Mohit Prabhushankar}
received his Ph.D. degree in electrical engineering from the Georgia Institute of Technology (Georgia Tech), Atlanta, Georgia, 30332, USA, in 2021. He is currently a Postdoctoral Research Fellow in the School of Electrical and Computer Engineering at the Georgia Institute of Technology in the Omni Lab for Intelligent Visual Engineering and Science (OLIVES). He is working in the fields of image processing, machine learning, active learning, healthcare, and robust and explainable AI. He is the recipient of the Best Paper award at ICIP 2019 and Top Viewed Special Session Paper Award at ICIP 2020. He is the recipient of the ECE Outstanding Graduate Teaching Award, the CSIP Research award, and of the Roger P Webb ECE Graduate Research Assistant Excellence award, all in 2022. He has delivered short courses and tutorials at IEEE IV'23, ICIP'23, BigData'23, WACV'24 and AAAI'24.
\end{IEEEbiography}

\begin{IEEEbiography}[{\includegraphics[width=1in,height=1.25in,clip,keepaspectratio]{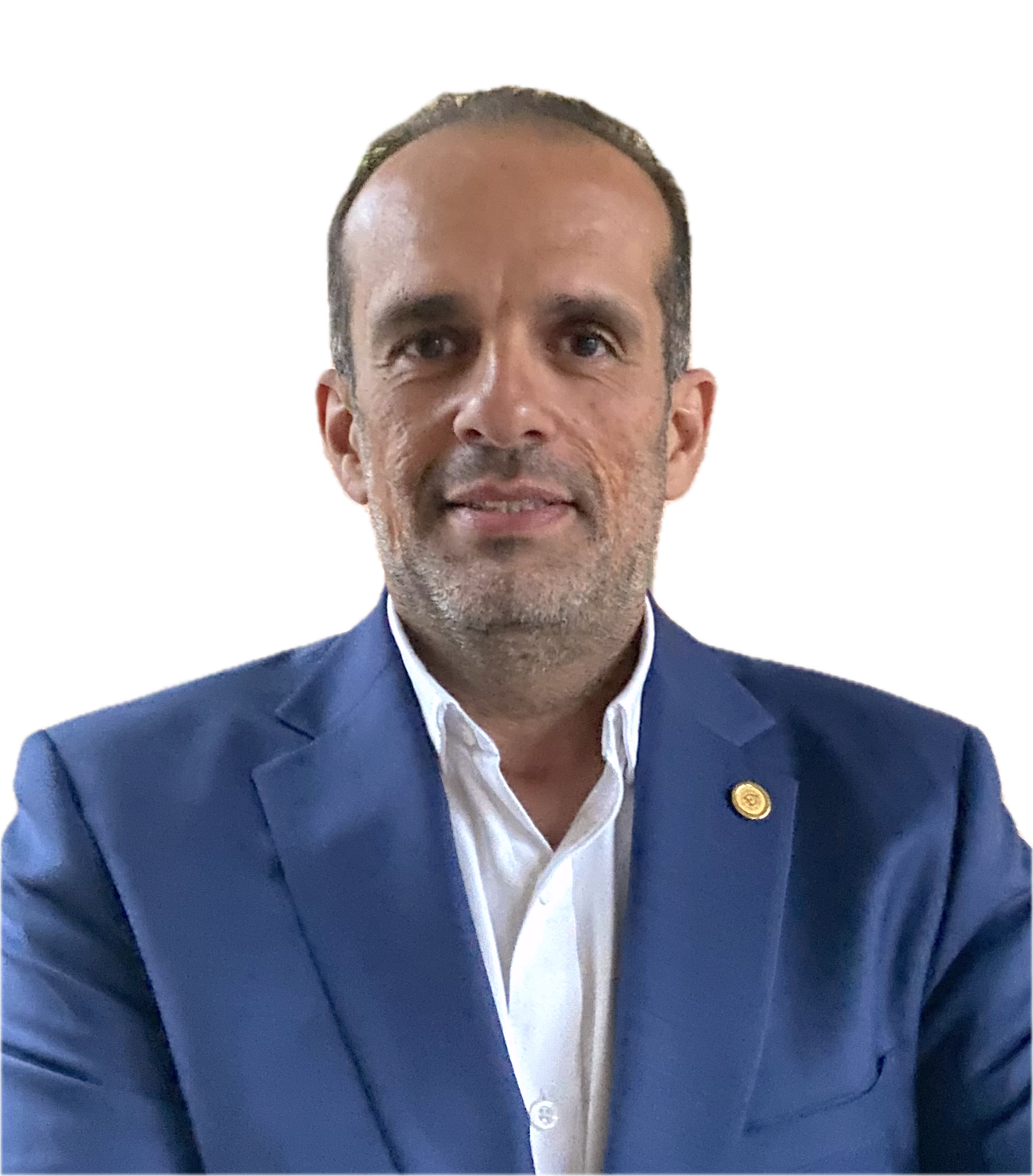}}]{Ghassan AlRegib}
is currently the John and Marilu McCarty Chair Professor in the School of Electrical and Computer Engineering at the Georgia Institute of Technology. In the Omni Lab for Intelligent Visual Engineering and Science (OLIVES), he and his group work on robust and interpretable machine learning algorithms, uncertainty and trust, and human in the loop algorithms. The group has demonstrated their work on a wide range of applications such as Autonomous Systems, Medical Imaging, and Subsurface Imaging. The group is interested in advancing the fundamentals as well as the deployment of such systems in real-world scenarios. He has been issued several U.S. patents and invention disclosures. He is a Fellow of the IEEE. Prof. AlRegib is active in the IEEE. He served on the editorial board of several transactions and served as the TPC Chair for ICIP 2020, ICIP 2024, and GlobalSIP 2014.  He was area editor for the IEEE Signal Processing Magazine. In 2008, he received the ECE Outstanding Junior Faculty Member Award. In 2017, he received the 2017 Denning Faculty Award for Global Engagement. He received the 2024 ECE Distinguished Faculty Achievement Award at Georgia Tech. He and his students received the Best Paper Award in ICIP 2019 and the 2023 EURASIP Best Paper Award for Image communication Journal. 
\end{IEEEbiography}



\vfill

\end{document}